\documentclass[aps,prl,twocolumn,superscriptaddress]{revtex4-1}%
\usepackage{amsfonts}
\usepackage{amsmath}
\usepackage{amssymb}
\usepackage{color}
\usepackage{graphicx}%
\usepackage{subfigure}

\begin{document}
\title{Doping dependence of the critical current in Bi$_2$Sr$_2$CaCu$_2$O$_{8+\delta}$}
\author{Muntaser Naamneh}
\affiliation{Department of Physics, Technion, Haifa 32000, Israel}
\author{Juan Carlos Campuzano}
\affiliation{Department of Physics, University of Illinois at Chicago, Chicago, IL 60607}
\author{Amit Kanigel}
\affiliation{Department of Physics, Technion, Haifa 32000, Israel}

\begin{abstract}
The doping dependence of the critical current
density, $J_{c}$, was measured in a series of high quality c-axis
oriented ${\rm Bi_2Sr_2CaCu_2O_{8+\delta}}$ thin films. For each
doping level we measured the temperature dependence of $J_{c}$.
We find that all samples have the same temperature dependence, but
the critical current at zero temperature, $J_{c}(0)$, has a
non-trivial doping dependence with a maximum at a unique doping
level in the overdoped side of the superconducting dome.

\end{abstract}
\maketitle

The normal state of the curpate High Tc superconductors (HTSC) and the origin of the enigmatic pseudo-gap (PG)  are still the focus of an intense debate. It is still not clear whether the PG is a precursor of superconductivity or a result of a competing order \cite{friend_or_foe}.  At low temperatures the samples are always superconducting; this prevents access to the real ground-state of the competing order phase, if indeed it exists. 
One possible way to overcome this problem is to use very high fields to suppress superconductivity and expose the competing phase. Experiments along this line were performed and indeed produced a lot of interesting new information \cite{CDW_field}, but one cannot rule out the possibility that the high magnetic field itself induces or stabilises the competing order.

A simple, but potentially powerful, approach for addressing this question would be to examine the dependence on doping of quantities associated with superconductivity, such as the critical current. In the present study, we report on systematic measurements of the critical current in high quality c-axis-oriented $\rm{Bi_2Sr_2CaCu_2O_{8+\delta}}$ (Bi2212) thin films by transport measurements.

%sample preparation
%Thin films of Bi2212 were deposited on (001)LaAlO$_3$ substrates using DC sputtering from a single target (Details of the sample preparation and characterization is given is the supplementary material).
%Several films were prepared with thicknesses of 100 nm to 300 nm as measured using an Atomic Force Microscope (AFM).

%After deposition, the doping level of the films can be controlled by annealing at different oxygen pressures and temperatures.  
%supp material

Thin films were prepared using DC sputtering from a single target. The target is a one inch pellet of sintered Bi2212, with the following composition: Bi$_{2.05}$Sr$_2$CaCu$_2$O$_y$. 
We add 5 \% excess Bi to the target since we noted that it results in better films. The films are grown on 0.5mm thick polished (001) LaAlO$_3$ substates. The substrate is heated to 860C and the sputtering 
is done in 3 Torr of O$_2$. The grow rate is about 100nm per hour. Several films were prepared with thicknesses of 100 nm to 300 nm as measured using an Atomic Force Microscope (AFM).

The composition of the films was measured using Energy Dispersive X-ray Spectroscopy (EDS) (Oxford instruments) in a SEM (FEI Quanta 200). We find very homogenous films with the composition Bi1.9Sr1.9Ca1.01Cu2Oy.
To change the doping of the films we used an annealing process. After the sputtering is finished the sample is cooled down to T$_{An}$ and then the pressure is lowered to about 100mTorr, the sample is held at T$_{An}$ for an hour and then cool down to room temperature. By changing T$_{An}$ we could cover most of the phase digram. 

\begin{figure}
[ptb]
\begin{center}
\includegraphics[trim=0cm 0cm 0cm 0cm,clip=true,width=9.5cm]{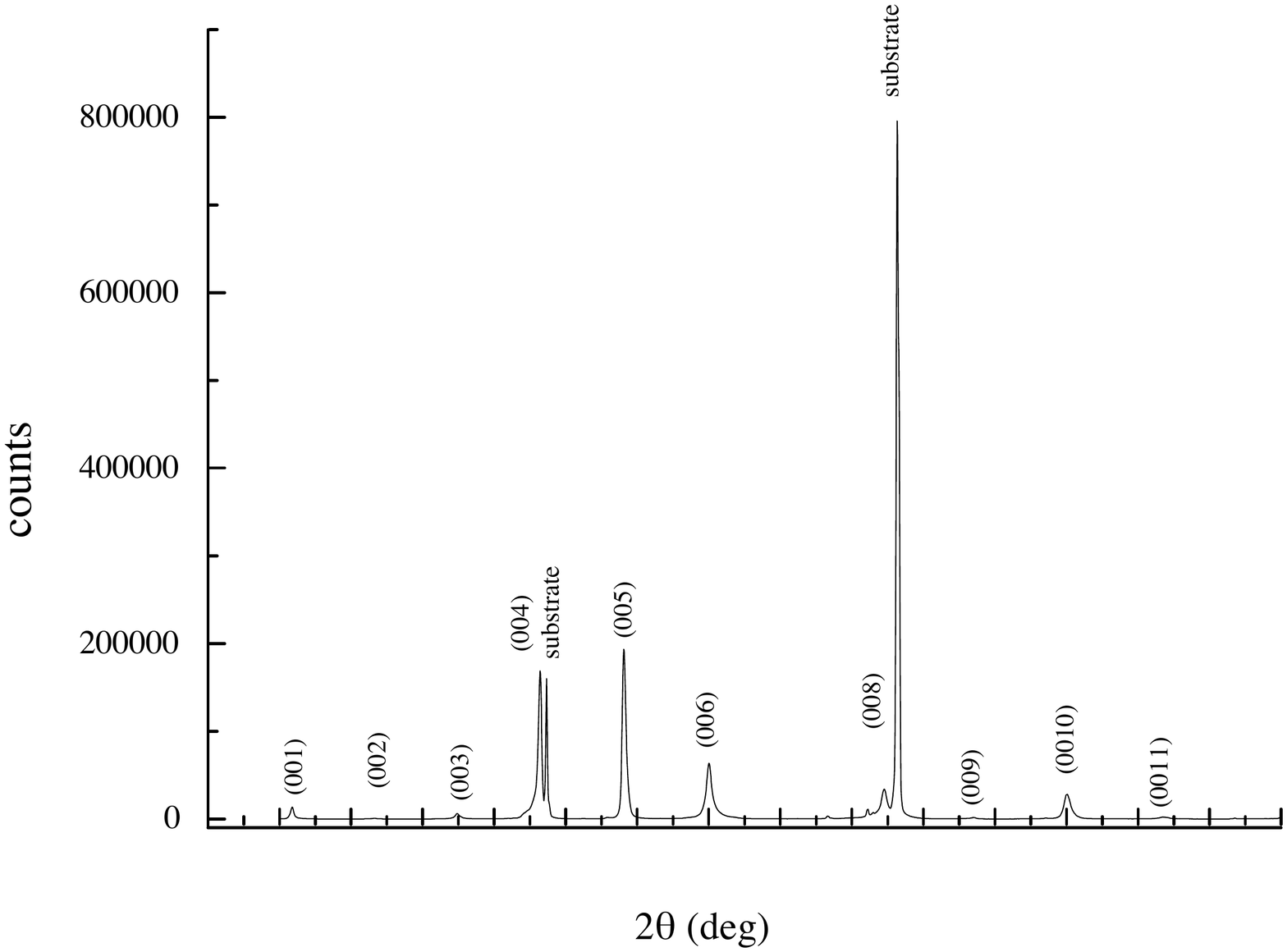}%
\caption{Typical XRD data for a film. Data collected with a Siemens D5000 diffractometer  using the Cu K$^{\alpha}$ line.}%
\label{xrd}%
\end{center}
\end{figure}

  To verify that the films grow epitaxially on the LAO substrates we performed X-ray diffraction (XRD) measurements, typical XRD data is shown in Fig. \ref{xrd}.  
We can identify clearly the (00n) series of peaks up to n=11, in addition we find peaks from the substrate. This is a clear indication that the films are c-axis oriented where the c-axis in perpendicular to the films plane. 

\begin{figure}
[ptb]
\begin{center}
\includegraphics[trim=0cm 0cm 0cm 0cm,clip=true,width=9.cm]{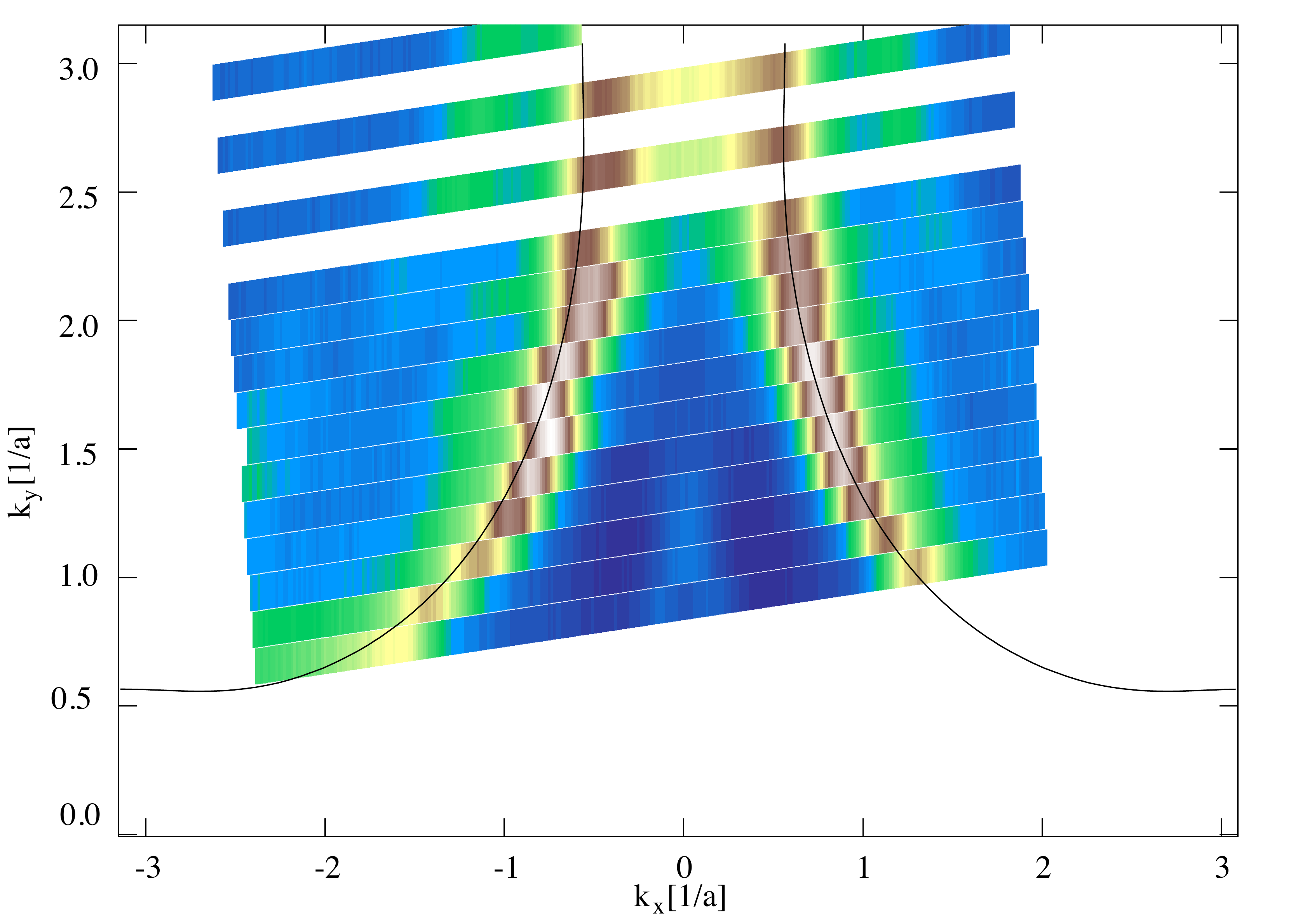}%
\caption{The Fermi surface of an optimally doped films as measured using ARPES. Experiment was done at the U1 beam-line at the SRC, Masidon, WI. The measurements were done using 22eV photons at T=40K. }%
\label{arpes}%
\end{center}
\end{figure}

Furthermore, to demonstrate  that the films are not only c-axis oriented but also epitaxially grown we show in Fig. \ref{arpes}  the Fermi-surface for an optimally doped film measured using Angle Resolved Photoemission Spectroscopy (ARPES).  ARPES is a very demanding technique in terms of the quality of the crystals used, there is no way to get ARPES data from polycrystalline samples. Overall the data obtained from films is very similar to the data obtained from single  crystals of Bi2212. We used the same films in several previous ARPES experiments \cite{PRL08,PRL12} . 

\begin{figure}
[ptb]
\begin{center}
\includegraphics[trim=0cm 0cm 0cm 0cm,clip=true,width=9.5cm]{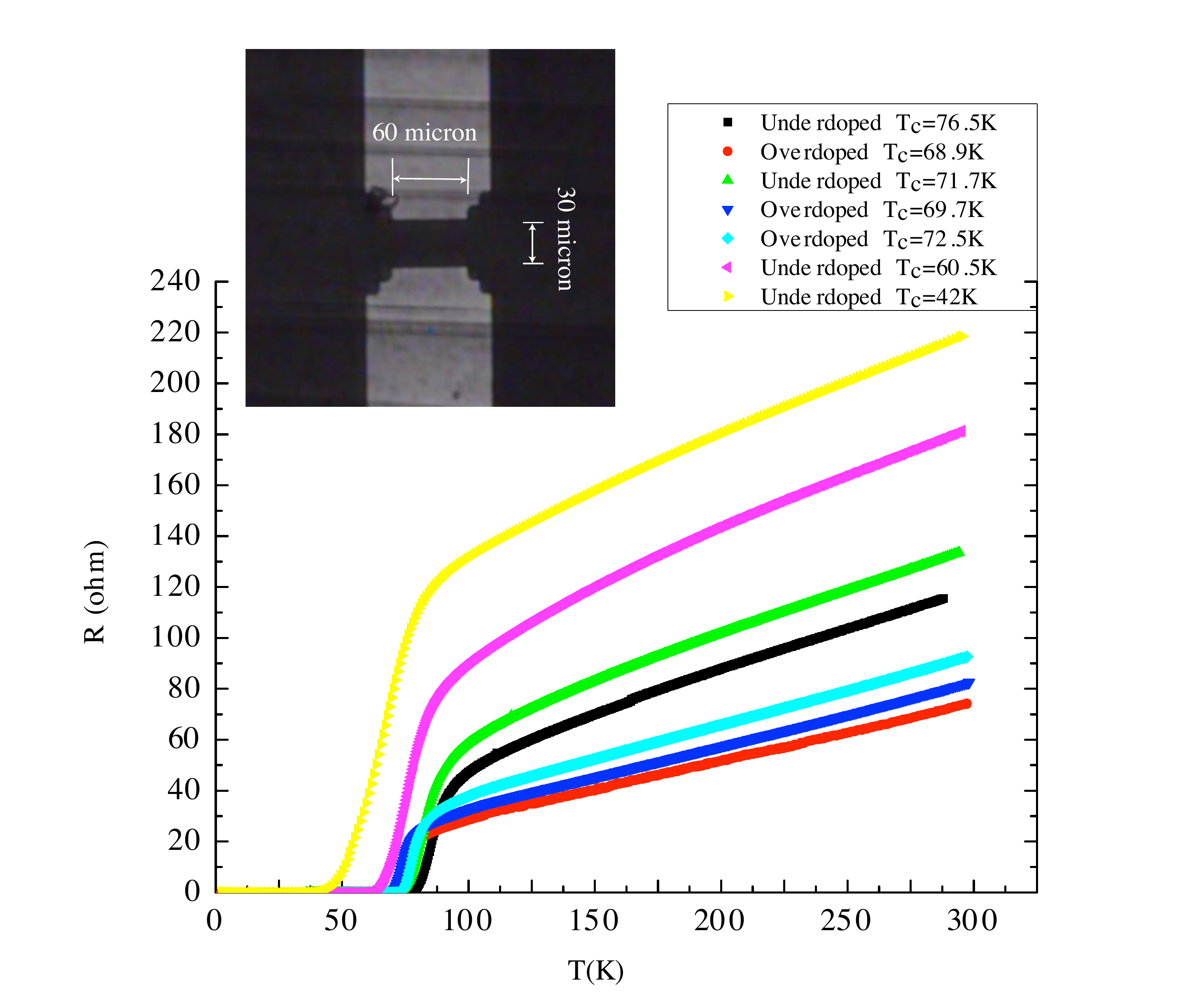}%
\caption{Resistance versus temperature for different different doping levels. Inset: Photograph of the Bi2212 bridge. The Au contacts are not shown in the figure.  }%
\label{R-Ts}%
\end{center}
\end{figure}

For running high current densities while keeping the overall current low, the films were photolithographically patterned into a $60 \mu$ long by $30 \mu$ bridge and etched in dilute $(\thicksim1\%)$ nitric acid $(HNO_3)$. An image of a typical bridge is shown in the inset of Fig.~\ref{R-Ts}. Gold contacts were then evaporated on the Bi2212 electrodes in order to reduce the contacts resistance.

%resistivity measurements

 The resistivity as a function of the temperature was measured using a standard four probe technique. Resistivity curves for several films are shown in Fig.~\ref{R-Ts}. From these measurements we  find the transition temperature of the films. We define T$_c$ as the temperature at which we measure zero resistivity. In addition, we can differentiate  underdoped films, which have "convex" curves at high temperatures, from overdoped films, which have "concave" curves in the normal state. The doping level of the films was estimated using the Presland formula \cite{Persland} $T_c=T_c^{max} (1- 82.6(p-0.16)^2)$ where $T_c$ is the critical temperature, $T_c^{max}$ is the critical temperature at optimal doping and $p$ is the number of holes per Cu atom. 
  
\begin{figure}
[ptb]
\begin{center}
\includegraphics[trim=0cm 0cm 0cm 0cm,clip=true,width=9cm]{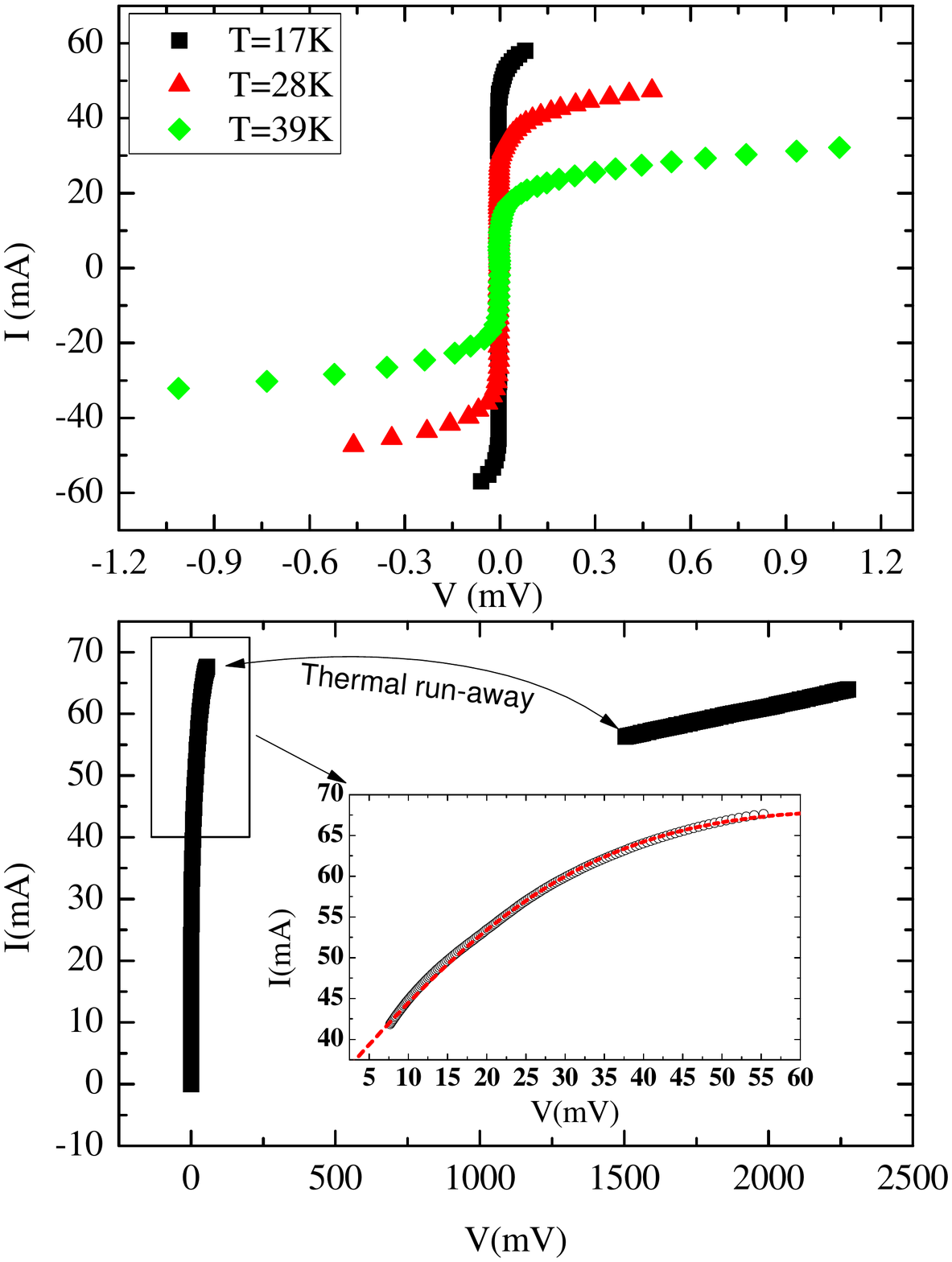}%
\caption{(upper panel) I-V curves at different temperatures for an under-doped sample with T$_c$=65K. (lower panel). A typical IV curve, showing the development of a voltage across the bridge and a flux-flow behaviour up to the LO instability jump followed by a thermal runaway into the normal state that is characterised by ohmic behaviour. The inset shows  a fit to the data in the flux-flow regime using Eq.~\ref{eqn:LOE} }%
\label{I-V}%
\end{center}
\end{figure}

%IV measurments
For each bridge, we  measured the voltage as a function of current (IV) at different temperatures, as shown in Fig.~\ref{I-V}. At each temperature, we ramped the current from zero up to 100 mA. The critical current was defined as the current that generates a voltage drop of $5 \mu V$ across the bridge. Increasing the current beyond this point induces flux flow and the resistance increases up to an instability point where the sample goes into another state and negative differential resistivity sets in, followed by a thermal runaway. This jump, which manifests itself as a distinct discontinuity in the IV characteristic, is a result of an electronic instability at a critical vortex velocity in the flux flow regime due to the shift of the quasiparticle distribution in the vortex core to higher energies. This mechanism for explaining the resistance in the flux flow regime was put forward by Larkin and Ovchinnikov (LO) \cite{LO} and was studied in detail in both conventional superconductors and the cuprates \cite{LOS}.   The LO theory gives

\begin{equation}
I-I_c=[\frac{V}{1+(\frac{V}{V^*})^2}+cV(1-\frac{T}{T_c})]\frac{1}{R_f}
\label{eqn:LOE}
\end{equation}
where $I_c$ is the critical current, $c$ is a number of order unity, $V^{*}$ is the critical voltage, and $R_f$ is a field-dependent resistance associated with the normal cores.  We fit Eq.~\ref{eqn:LOE}, where $I_c,c,R_f$ are free parameters, to our IV data in the flux flow regime near the instability point. The results of this fit are shown in Fig.~\ref{I-V}.

The typical critical current densities we measure are 1-10$\times $10$^6$ A/cm$^2$. These values definitely correspond to the de-pinning critical current and are lower by at least an order of magnitude than the estimated de-pairing currents in the cuprates \cite{Pulse_YBCO}. The good agreement of our data with the predictions of the LO theory also indicates that the critical current we measure is the minimal current at which vortex motion develops.  

%same film re-oxygen

\begin{figure}
[ptb]
\begin{center}
\includegraphics[trim=0cm 0cm 0cm 0cm,clip=true,width=9cm]{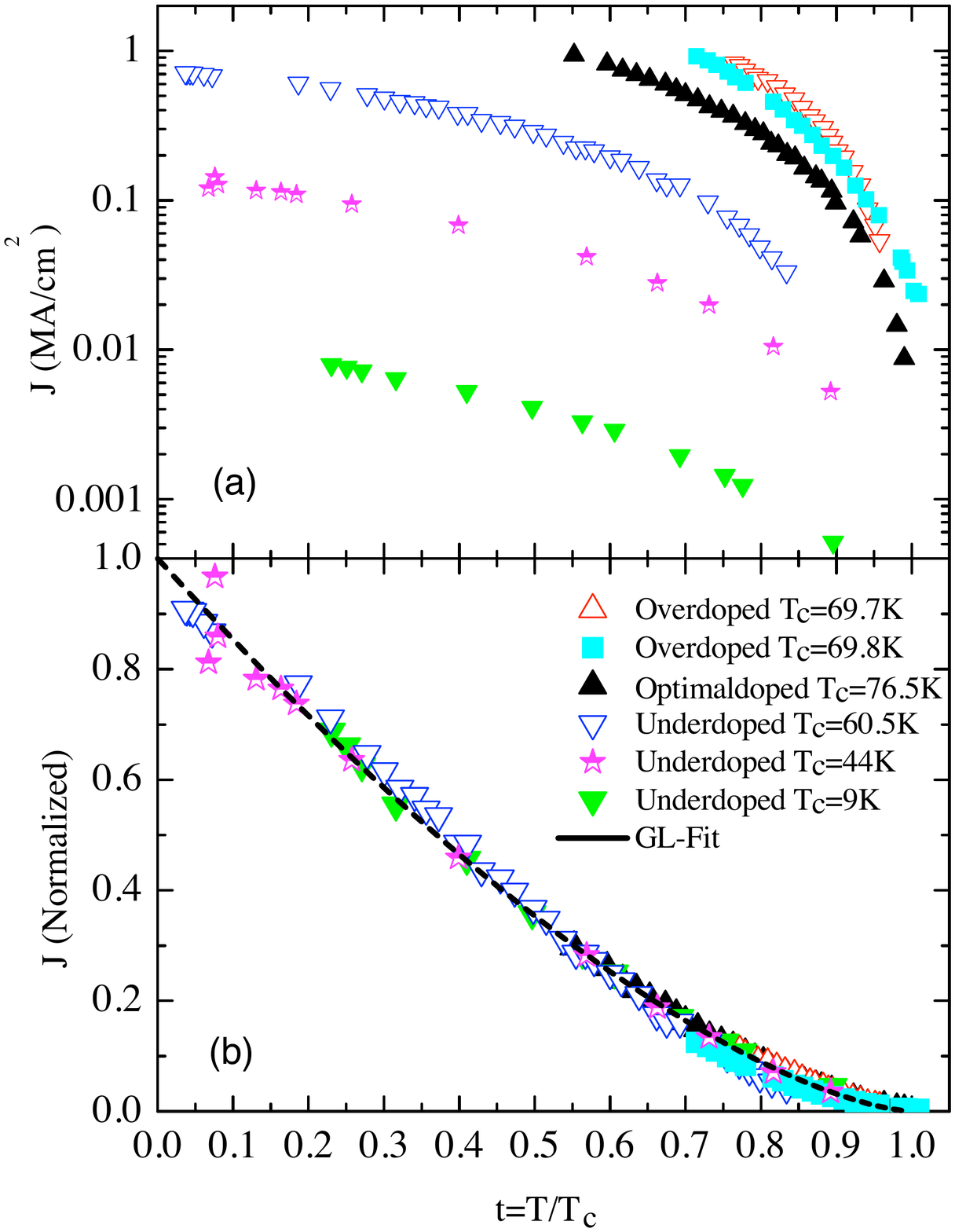}%
\caption{(a) The critical current density vs. the reduced temperature, $t=T/T_c$, for different doping levels on a log-normal scale. (b) The critical current density normalised by the value at zero temperature as a function of the reduced temperature. The dashed line shows $(1-t)^{3/2}$ }.
\label{J-t}%
\end{center}
\end{figure}

The measured critical current density can depend on the bridge dimensionality, surface morphology, and the pinning forces associated with various structural defects. Efficient pinning centres are formed by extended growth defects, such as twining and grain boundaries, and by smaller point-like defects \cite{defects}. These extended defects are formed during the growth of the film and may differ from one film to another. In order to exclude the possibility that different bridges have different concentrations of structural defects, we change the doping level of the {\it same} film by re-annealing at different temperatures and oxygen pressures. For each doping level, the critical current was measured from the IV characteristic at various temperatures.  In Fig.~\ref{J-t}(a) we plot the critical current as a function of the reduced temperature ($\frac{T}{T_c}$) for various doping levels. The temperature dependence of the critical current density follows fairly well a power law behaviour, which can be described by

\begin{equation}
J(T)=J_0(1-\frac{T}{T_c})^{\frac{3}{2}}
\label{eqn:GL}
\end{equation}

In Fig.~\ref{J-t}(b) we demonstrate that the power law temperature dependence does not depend on the doping level over the entire doping range. When we normalise the critical current using the zero temperature value, J$_c$(0), and we plot it against the reduced temperature for different doping levels we get a perfect collapse of all the data onto a single curve.   The value of J$_c$(0) was obtained by fitting the J-T data using Eq.~\ref{eqn:GL}. The temperature dependence we find is in agreement with the results of the Ginzburg-Landau (GL)  theory for the de-pairing current \cite{tinkham}. Similar temperature dependence was found in pulsed-current experiments done on very thin films of YBCO \cite{Pulse_YBCO}.

\begin{figure}
[ptb]
\begin{center}
\includegraphics[trim=0cm 0cm 0cm 0cm,clip=true,width=9cm]{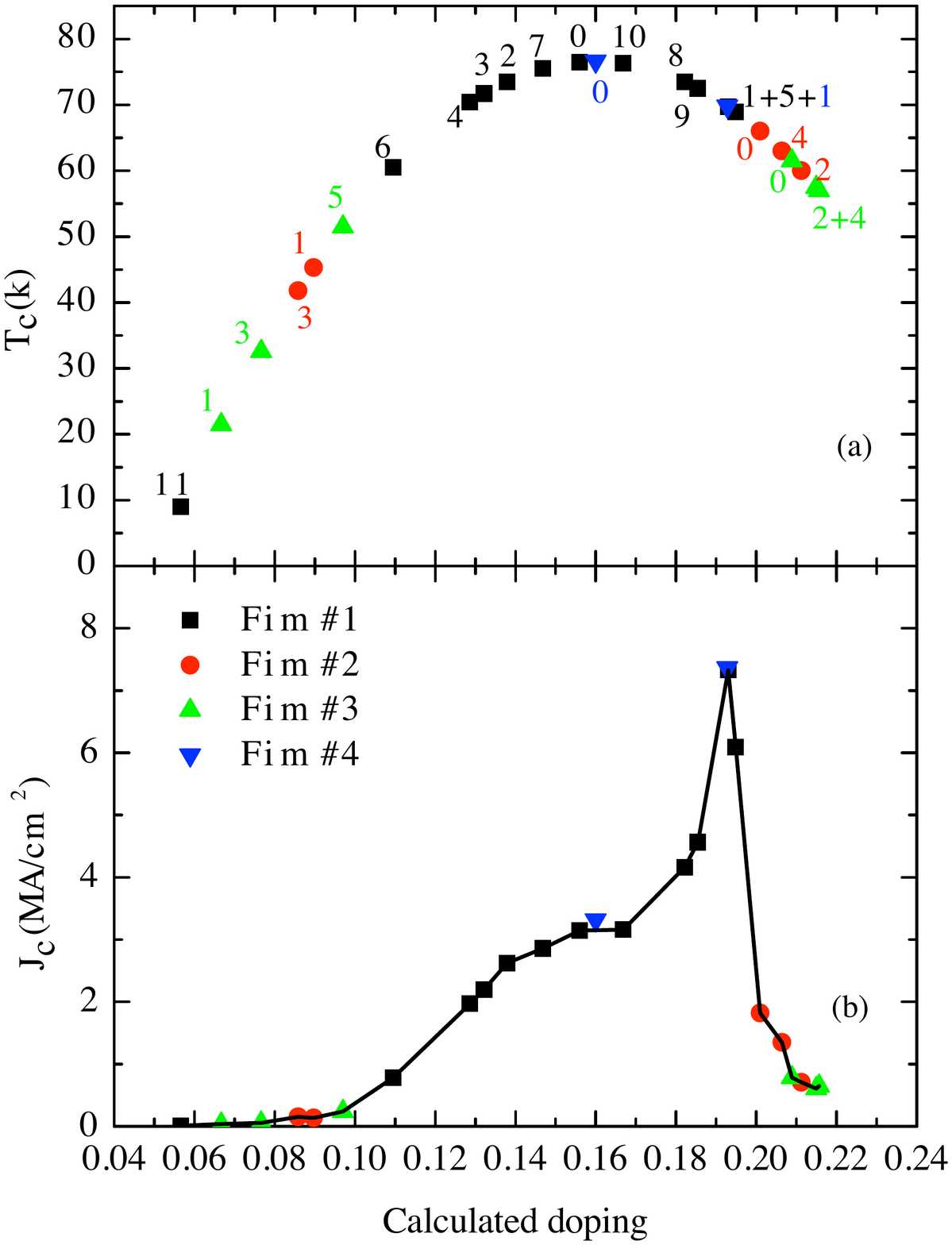}%
\caption{(a) T$_c$ of all the measured samples plotted as function of the calculated doping (using the Presland formula). The different symbols represent different bridges and the numbers show the sequence of re-annealing for the same bridge. (b)  The zero temperature critical current density as a function of the calculated doping for all the samples shown in panel (a).}%
\label{J-p}%
\end{center}
\end{figure}

In Fig. \ref{J-p} we show the doping dependence of the critical current at zero temperature, J$_c$(0). The plot is constructed from data taken using four different films that were re-annealed several times. In the same figure we show the transition temperature of all the measured samples. The different colors represent different films and the numbers show the ordering of the annealing sequence. 
Several interesting features can be seen in Fig. \ref{J-p}. First, the critical current does not simply follow T$_c$, or the doping. For very 
under-doped samples (p$\lesssim$0.1) the critical current is very low and grows more slowly than T$_c$ with doping. Second, there is a pronounced peak in the critical current at p $\approx$ 0.19. Beyond that doping level, the critical current decreases sharply with doping.
The non-trivial doping dependence of the critical current density is our main result; in the rest of the paper we are trying to derive some insight from the data making simple assumptions.

The typical value we find for the critical current density and the fact that there is an extended range of current density beyond the critical value for which we find flux-flow indicate that we are measuring the minimal current for de-pinning of flux vortices.  The motion of a single flux line is governed by three forces: the force acting on the vortex due to the current flow, the pinning force and elastic forces \cite{Blatter}. When we consider finite magnetic fields the vortex-vortex interaction should be also taken into account. Vortex physics can be very complicated; the specific conditions in our case allow for some simplification.

All our measurements are done in  a zero external field; the self-field is estimated to be less than 20 Gauss \cite{Zeldov_Jc}, which leads us to ignore interaction effects. Since we are dealing with thin films, it is natural to assume that elasticity does not play a significant role \cite{Ex-defects1}. 
We are left with pinning forces; the critical current is the current for which the Lorentz force density is equal to the pinning force density

\begin{equation}
f_L=J_c B = f_p 
\end{equation}

The pinning force density is given by the gradient of the pinning energy divided by the volume per vortex \cite{CPM}:
\begin{equation}
f_p=\nabla U_p / V_{\phi}
\end{equation}
For low fields we assume that the volume per vortex is $V=da^2$, where $a$ is the average separation between vortices and $d$ the film thickness. In its simplest form the gradient of the pinning force is given by $\eta \nabla U_C \simeq \eta U_C/\xi$, where $U_C$ is the condensation energy, $\xi$ the GL coherence length and $\eta$ a number of order one to take into account partial suppression of the order parameter in the pinning site. Then we get for the pinning force density \cite{CPM} 
\begin{equation}
f_p=J_cB \simeq \eta \frac{1}{da^2}\frac{1}{\xi}U_C(\pi d \xi^2)
\end{equation}

and for the critical current density

\begin{equation}
J_c \simeq \frac{\eta \pi}{\phi_0} \xi U_C 
\end{equation}

We get simply that the critical current is given by the condensation energy times the coherence length. It is interesting to note that according to the GL theory the temperature dependence of $\xi U_C$  is (1-T/T$_c$)$^{3/2}$, in good agreement with our data.

Based on this simple model the doping dependence of the critical current is governed mainly by the doping dependence of the condensation energy since $\xi(p)$  decreases linearly with doping \cite{Ong}. The condensation energy is defined as the difference between the free energy of the superconductor and that of the normal metal at the same thermodynamic conditions. In a case where a different order is competing with SC, the condensation energy should be thought of as the energy difference between the SC state and a state where the competing order is stabilised due to the reduction of the SC order. It is well known that the electronic and magnetic structure of the vortex core in the cuprates is different than found in the normal state \cite{Lake, Hoffman}. 

 In such a situation, the doping dependence of the condensation energy does not reflect only the evolution of SC with doping but also the evolution of the competing order with doping. In the BCS theory the condensation energy is given by 0.5N(0)$\Delta^2$(0), where N(0) is the density of states at the Fermi-level and $\Delta$(0) is the superconducting energy gap at zero temperature. The anti-nodal gap in Bi2212 is known to decrease linearly with doping following the T$^{*}$ line \cite{Utpal}, suggesting a completely different doping dependence of the critical current as compared to that shown by our results. A somehow better agreement with BCS can be observed if we use the SC gap calculated using the gap slope around the node \cite{Vishik}. In any case, the sharp feature around x=0.19 is hard to explain unless we assume that it reflects the doping dependence of the competing order. It was suggested previously that x=0.19 marks the end of the competing phase \cite{x0p19}. 
 
 Our results are very similar to those reported by Tallon {\it et al.} \cite{Tallon_Jc_YBCO}, with the same clear peak at x=0.19. In this work aligned powders of YBCO were used and the critical current was extracted from magnetisation measurements, very different conditions compared to our experiment; nevertheless, the results are basically identical. This suggests that the doping dependence is governed by some basic property of the material and is not too sensitive to other aspects of vortex physics.
 
 A completely different approach for understanding the doping dependence of the critical current was presented by Goren and Altman \cite{Lilach}. Using a variational method the authors showed that the way current destroys SC changes with doping. While for overdoped samples the current destroys the gap in the usual BCS way, for underdoped samples the current creates a resistive state by reducing the superfluid stiffness without closing the gap.  The transition between these two mechanisms produces a maximum in the critical current in the overdoped side slightly above optimal doping \cite{Lilach}.   

To summarize, we measured the doping the dependence of the critical current density in Bi2212 films. We found that the critical current increases with doping, reaching a sharp maximum at a doping level of 0.19 holes per Cu atom and decreases beyond that doping level. We suggest that this doping dependence reflects that of the condensation energy.  The condensation energy in these SC can be sensitive to not only  the evolution of SC with doping but also  the doping dependence of a competing order that is believed by many to exist in the under-doped side of the phase diagram of the cuprates up to a doping level of about 0.19.

We thank O. Auslaender , Y. Yeshurun, L. Goren and E. Altman for useful discussions.  
This research was supported by Grant No 2010313 from the United States-Israel Binational Science Foundation (BSF). The Synchrotron Radiation Center is supported by NSF DMR- 0084402.


\begin{thebibliography}{99}

\bibitem{friend_or_foe} M.R. Norman, D. Pines and C. Kallin, Adv. Phys. \textbf{54}, 715 (2005).

\bibitem{CDW_field} Tao Wu, Hadrien Mayaffre, Steffen Krämer, Mladen Horvatić, Claude Berthier, W. N. Hardy, Ruixing Liang, D. A. Bonn and Marc-Henri Julien, Nature \textbf{477}, 191 (2011).

\bibitem{PRL08} A. Kanigel, U. Chatterjee, M. Randeria, M. R. Norman, G. Koren, K. Kadowaki, and J. C. Campuzano,  Phys. Rev. Lett. \textbf{101}, 137002  (2008).

\bibitem{PRL12} Y. Lubashevsky, A. Garg, Y. Sassa, M. Shi, and A. Kanigel, Phys. Rev. Lett. \textbf{106},  047002 (2011).


\bibitem {Persland} M.R. Presland, J.L. Tallon, R.G. Buckley, R.S. Liu and N.E. Flower , Physica C. \textbf{176},95 (1991).

\bibitem {LO} A. I. Larkin and Yu.N. Ovchinnikov, Zh. Eksp. Teor. Fiz. \textbf{68},
1915 (1975) [Sov. Phys. JETP. \textbf{41}, 960 (1976)].

\bibitem {LOS}R. P. Huebener, Magnetic Flux Structures in Superconductors
(Springer, Berlin, 2001), 2nd ed.

 \bibitem{Pulse_YBCO} W. Langa, I. Puicaa, K. Sirajb, M. Peruzzib, J.D. Pedarnigb and D. Bäuerleb, Physica C \textbf{460-462}, 827 (2007).

\bibitem {defects} B. Dam, J. M. Huijbregtse, F. C. Klaassen, R. C. F. van der Geest, G. Doornbos, J. H. Rector, A. M. Testa, S. Freisem, J. C. Martinez, B. Stäuble-Pümpin and R. Griessen , Nature \textbf{399},439 (1999).

\bibitem{tinkham} M. Tinkham, Introduction to superconductivity, second ed., McGraw-Hill, New York, NY, 1996.



\bibitem{Blatter} G. Blatter, M. V. Feigel'man, V. B. Geshkenbein, A. I. Larkin, and V. M. Vinokur, Rev. Mod. Phys. \textbf{66}, 1125 (1994).



\bibitem{Zeldov_Jc} E. Zeldov, John R. Clem, M. McElfresh, and M. Darwin, Phys. Rev. B \textbf{49}, 9802 (1993).


\bibitem {Ex-defects1} T.L. Hylton, M.R. Beasley,  Phys. Rev. B \textbf{41}, 11669�11672 (1990). 


\bibitem {CPM} J. G. Ossandon, J. R. Thompson, D. K. Christen, B. C. Sales, H. R. Kerchner, J. O. Thomson, Y. R. Sun, K. W. Lay, and J. E. Tkaczyk, Phys. Rev. B \textbf{45}, 12534 (1992).

\bibitem {Ong} Yayu Wang, S. Ono, Y. Onose, G. Gu, Yoichi Ando, Y. Tokura, S. Uchida, and N. P. Ong, Science \textbf{299}, 86 (2003).



\bibitem{Lake} B. Lake, H. M. Rønnow, N. B. Christensen, G. Aeppli, K. Lefmann, D. F. McMorrow, P. Vorderwisch, P. Smeibidl, N. Mangkorntong, T. Sasagawa, M. Nohara, H. Takagi and T. E. Mason \emph{et al.}, Nature \textbf{415}, 299 (2002).

\bibitem{Hoffman} J. E. Hoffman, E. W. Hudson, K. M. Lang, V. Madhavan, H. Eisaki, S. Uchida and J. C. Davis, Science, \textbf{295}, 466 (2002).


\bibitem{Utpal}U. Chatterjee, M. Shi, D. Ai, J. Zhao, A. Kanigel, S. Rosenkranz, H. Raffy, Z. Z. Li, K. Kadowaki, D. G. Hinks, Z. J. Xu, J. S. Wen, G. Gu, C. T. Lin, H. Claus, M. R. Norman, M. Randeria and J. C. Campuzano, Nature Phys. \textbf{6}, 99 (2010).

\bibitem{Vishik} I. M. Vishik, M. Hashimoto, Rui-Hua He, Wei-Sheng Lee, Felix Schmitt, Donghui Lu, R. G. Moore, C. Zhang, W. Meevasana, T. Sasagawa, S. Uchida, Kazuhiro Fujita, S. Ishida, M. Ishikado, Yoshiyuki Yoshida, Hiroshi Eisaki, Zahid Hussain, Thomas P. Devereaux and Zhi-Xun Shen, PNAS \textbf{109}, 18332 (2012). 

\bibitem{x0p19}  K. Fujita, Chung Koo Kim, Inhee Lee, Jinho Lee, M. H. Hamidian, I. A. Firmo, S. Mukhopadhyay, H. Eisaki, S. Uchida, M. J. Lawler, E.-A. Kim and J. C. Davis, Science \textbf{344}, 612 (2014); N. E. Hussey, R. A. Cooper, Xiaofeng Xu, Y. Wang, I. Mouzopoulou, B. Vignolle and C. Proust, Phil. Trans. R. Soc. A \textbf{369}, 1626 (2011); J.W. Loram, K.A. Mirza, J.M. Wade, J.R. Cooper, W.Y. Liang, Physica C \textbf{235-240}, 134 (1994); J. L. Tallon, J. W. Loram, C. Panagopoulos, J. Low. Temp. Phys. \textbf{131}, 387 (2003).

\bibitem{Tallon_Jc_YBCO} J.L. Tallon, G.V.M. Williams and J.W. Loram, Physica C \textbf{338}, 9 (2000).
\bibitem{Lilach} Lilach Goren and Ehud Altman, Phys. Rev. Lett. \textbf{104}, 257002 (2010).

\end{thebibliography}
\end{document}